\documentclass[9pt,conference]{IEEEtran}
\IEEEoverridecommandlockouts

\usepackage{epsfig}
\usepackage{graphicx}
\usepackage{amsmath}
\usepackage{amssymb}
\usepackage{ifpdf}
\usepackage{cases}
\usepackage[pagebackref=true,breaklinks=true,letterpaper=true,colorlinks,bookmarks=false]{hyperref}
\usepackage{subfigure}

\def\BibTeX{{\rm B\kern-.05em{\sc i\kern-.025em b}\kern-.08em
    T\kern-.1667em\lower.7ex\hbox{E}\kern-.125emX}}

\newcommand{\eR}{\mathbb{R}}

\begin{document}
\title{A Fully Convolutional Network for MR Fingerprinting
\thanks{This work is funded by the ERC Advanced grant 694888, C-SENSE.}
}


\author{\IEEEauthorblockN{Dongdong Chen$^{\dag}$, \emph{\emph{ Mohammad Golbabaee}}$^{\ddag}$, \emph{\emph{ Pedro A. G\'{o}mez}}$^{\circ,\star}$,  \emph{\emph{ Marion I. Menzel}}$^{\star}$,  \emph{\emph{ Mike E. Davies}}$^{\dag}$ }
\IEEEauthorblockA{$^{\dag}$ \emph{\emph{School of Engineering, The University of Edinburgh}}, $^{\ddag}$ \emph{\emph{Computer Science department, University of Bath}},
\\ $^{\circ}$ \emph{\emph{School of Bioengineering, Technische Universit\"{a}t M\"{u}nchen}}, $^{\star}$ \emph{\emph{GE Healthcare}}.}}

\maketitle

\begin{abstract}
Magnetic Resonance Fingerprinting (MRF) methods typically rely on dictionary matching to map the temporal MRF signals to quantitative tissue parameters. These methods suffer from heavy storage and computation requirements as the dictionary size grows. To address these issues, we proposed an end to end fully convolutional neural network for MRF reconstruction (MRF-FCNN), which firstly employ linear dimensionality reduction and then use neural network to project the data into the tissue parameters manifold space. Experiments on the MAGIC data demonstrate the effectiveness of the method.
\end{abstract}


\section{Introduction}
Magnetic Resonance Fingerprinting (MRF) \cite{ma2013magnetic} has emerged as a promising quantitative Magnetic Resonance Imaging approach, which can significantly reduce the acquisition time needed for quantitative measurements. However, the conventional MRF methods suffer from the heavy storage and computation requirements of a dictionary-matching (DM)  step due to the growing size and complexity of the fingerprint dictionaries in multi-parametric quantitative MRI applications \cite{davies2014compressed}. Recently, deep neural networks have been used for MRF reconstruction \cite{hoppe2017deep,cohen2018mr}.
Thanks to the powerful ability of neural networks to approximate nonlinear functions and solving many learning problems \cite{chen2017unsupervised,chen2018learning,li2019dual,chen2018slrbm}, these methods propose  to exploit deep neural networks to replace the dictionary and the lookup-table used in conventional MRF reconstruction approaches. However, these approaches typically relied on fully-sampled instead of typically available sub-sampled k-space data \cite{song2019hydra}. Alternately, by imposing a linear dimension reduction procedure, our proposed MRF-Net is able to accurately approximate the DM step saving more than 60 times in memory and computations \cite{mo2019geometry}. This paper extends the learning capability of the MRF-Net by including a fully convolutional architecture that is capable of capturing both spatial and temporal structures.

\section{MRF-FCNN}
The proposed MRF-FCNN consists of two components: a linear projector $\mathcal{P}_0: \eR^{m\times d_0} \rightarrow \eR^{m\times d_1}$ and a neural network projector $\mathcal{P}_1: \eR^{m\times d_1} \rightarrow \eR^{m\times d_2}$, where $m$ is the number of voxels, $d_0$ is the number of acquired time points (i.e., dimensionality of the fingerprinting), $d_1=10$ is the reduced dimensionality, and $d_2=3$ corresponding to the desired tissue's intrinsic parameters $\Theta = [T1, T2, PD]$. The former aims to learn a linear projection onto the subspace of clean fingerprints, the target of the latter is to nonlinearly project the dimension-reduced data onto the manifold of $\Theta$. Therefore, MRF-FCNN finally approximates the following transformation $\mathcal{F}$:

\begin{equation}
    \mathcal{F}: \mathcal{P}_0 \circ \mathcal{P}_1
\end{equation}

\begin{figure*}[t]
\begin{center}
\includegraphics[width=0.91\linewidth]{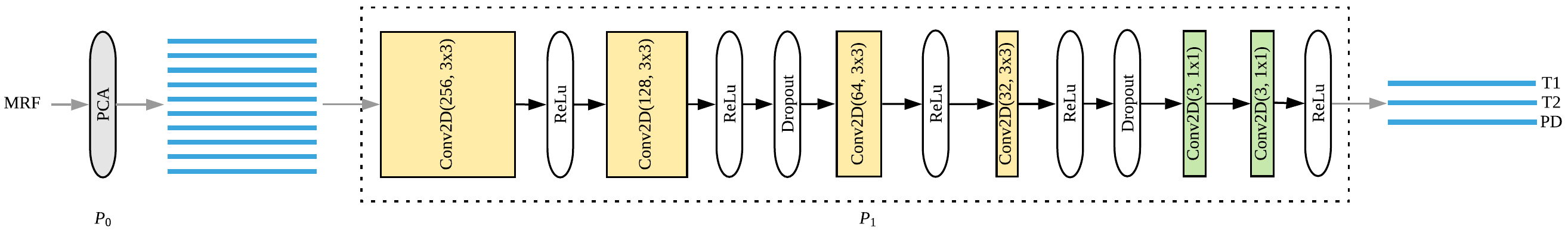}
\end{center}
\caption{An illustration of the proposed MRF-FCNN. Inputs are the voxel sequences and output are the per-voxel T1, T2 and PD parameters.}
\label{fig:1}
\end{figure*}

In this work, we apply principal component analysis (PCA) as the $\mathcal{P}_0$ and a concisely designed fully convolutional neural network as the MRF-FCNN is summarized in  Figure \ref{fig:1}. It starts with an unsupervised learning layer (gray) which learns a linear projection onto the subspace of clean fingerprints through PCA, then keep $\mathcal{P}_0$ fixed during the training of the other layers. The main component (dotted box) of the MRF-FCNN is designed with stacks of separable convolutional layers (yellow) with kernel size $3\times 3$, and decreasing feature maps $(256, 128, 64, 32)$ for fine texture features learning, and finally ends with two convolutional layers (green) with the same kernel size $1\times 1$ and $3$ feature maps for each layer. The ReLu are used as the activation function, as it provides a piece-wise affine approximation to the Bloch  response manifold projection  \cite{mo2019geometry} (i.e. a transformation from tissue's intrinsic parameters $[T1, T2]$ to it's corresponding temporal signature), the followed dropouts are included to prevent over-fitting. Our empirical studies show that 1) not using pooling layers 2) and including the $1\times 1$ convolutional layers at the end of the model are crucial to the reconstruction, as they help to prevent local blurring in the reconstruction.


Our approach performs a sharp and accurate parameter estimation. The proposed MRF-FCNN uses spiral sub-sampled data but it reconstructs the data with similar accuracy to the Cartesian sampled images acquired using a specific protocol, e.g. MAGIC \cite{magic2015}. In addition, benefit from the dimensionality-reduction operator, the MRF-FCNN requires far less units and training resource which distinguishes from other mainstream deep learning approached applied to MRF  \cite{hoppe2017deep,cohen2018mr}, and our experimental results show that the MRF-FCNN does not suffer from common blurring artifacts in spiral sampling protocols.





\section{Experiment}

\begin{figure*}[t]
\begin{center}
    \subfigure[]{\includegraphics[width=0.33\linewidth]{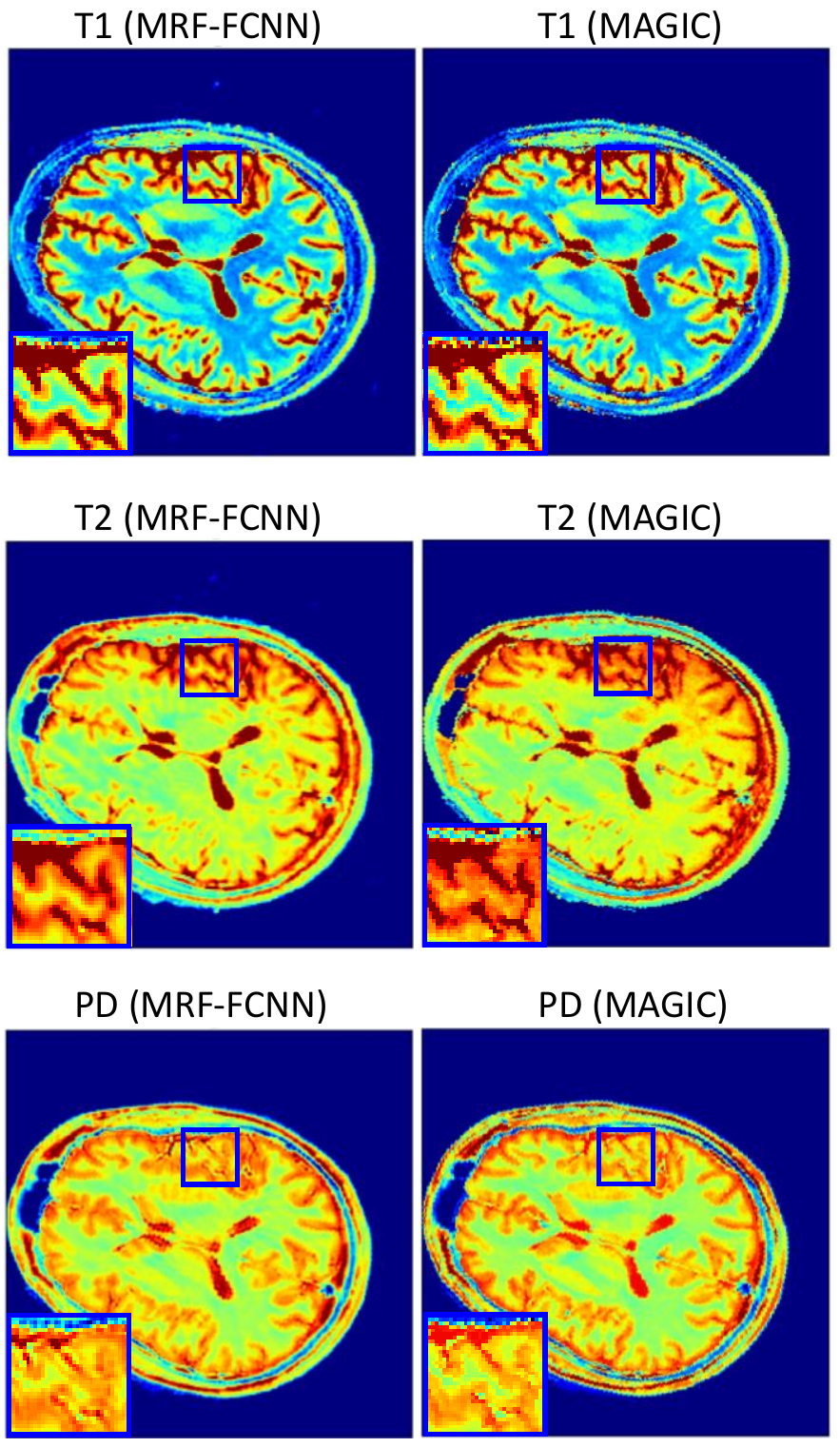}
    \label{fig:1a} }
    \hspace{6ex}
    \subfigure[]{\includegraphics[width=0.33\linewidth]{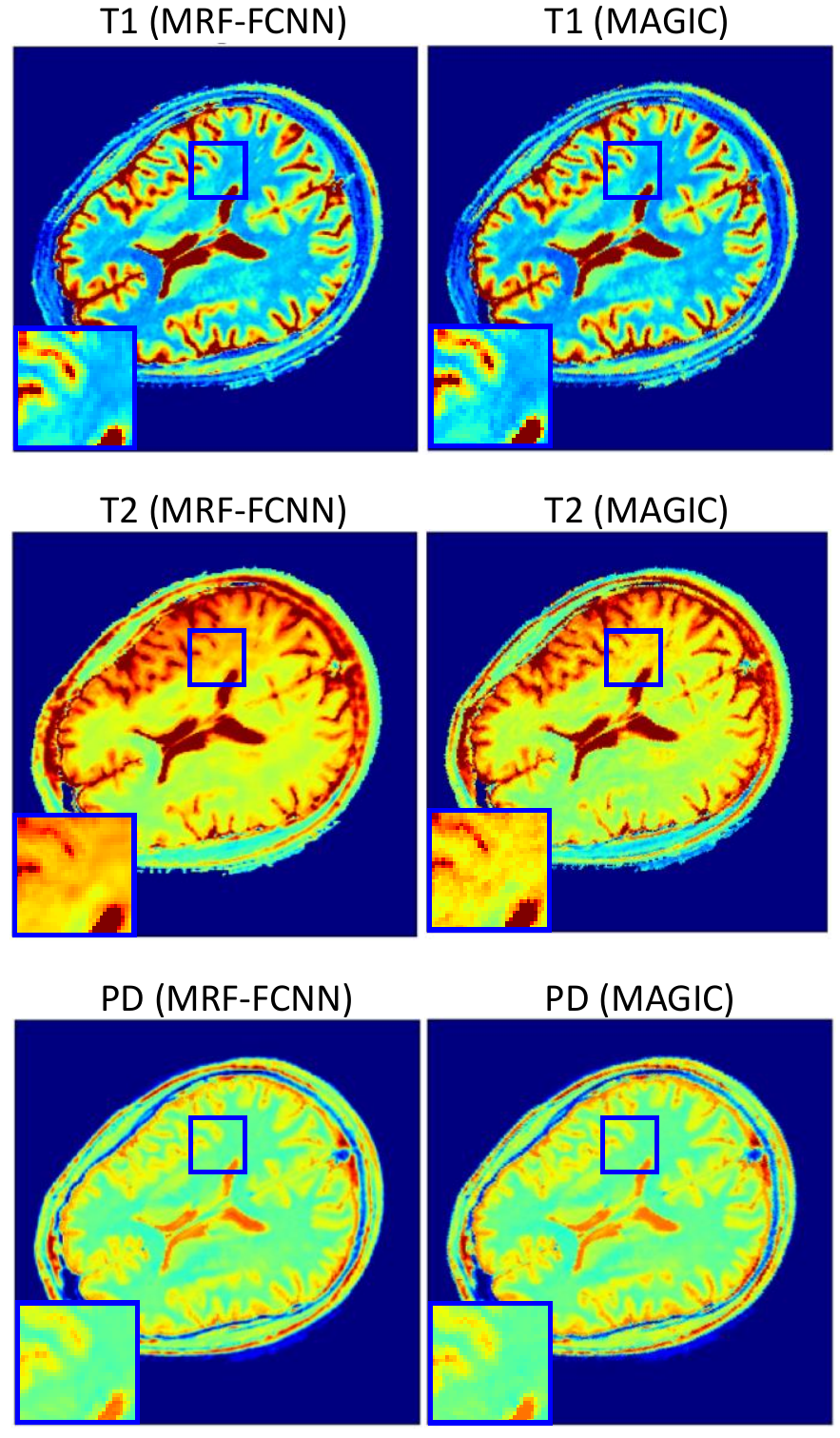}
    \label{fig:1b} }
\end{center}
\caption{Two examples of the reconstructed T1, T2 and PD maps using the proposed MRF-FCNN.  For each example (a) and (b), the left column of images are our MRF reconstruction results using FISP protocol and spiral sampling, the right images are the ground truth parametric maps acquired using MAGIC protocol with Cartesian sampling. Results indicate we can get high-quality reconstruction similar to MAGIC without suffering from blurring artifacts.
}
\label{fig:2}
\end{figure*}

We test the MRF-FCNN on a simulated human brain MRF data. To be specific, the ground truth used for this simulation were acquired using rather longer protocol MAGIC \cite{magic2015}.
In this work, we set the image scale $m=256\times 256$, and we collect ground truth (GT) parametric maps from 8 volunteers (20 brain slices each) using MAGIC quantitative MRI protocol with Cartesian sampling.
These parametric maps have been used for simulating MRF acquisition using the  Fast imaging with Steady State Precession (FISP) \cite{jiang2015mr} protocol and spiral sampling. Accordingly, the input to MRF-FCNN is the MRF measurements and output are the GT parametric maps. To avoid overfitting, the standard data augmentation is conducted by adding the translation, rotation, scale, and noise. The MRF reconstruction results (Figure \ref{fig:2}) shows the proposed MRF-FCNN could generate high-quality reconstruction similar to MAGIC without suffering from blurring artifacts. More importantly, the standard dictionary matching (search) approaches would typically take a couple of minutes for reconstruction, but MRF-FCNN only takes around 0.29 seconds for a single slice MRF reconstruction. This means we can use the MRF-FCNN framework and get similar quality for quantifying tissues as MAGIC but in much shorter acquisition time.  Detailed comparison with other methods in terms of reconstruction quality, computational performance and applicability to real-world data will be addressed in future work.

\bibliographystyle{IEEEtran}
\bibliography{dd}
\end{document}